\documentclass{aa}  
\usepackage{graphicx}
\usepackage{txfonts}
\usepackage{lipsum}
\usepackage{subcaption}
\usepackage{lscape}
\usepackage{placeins}
\usepackage[colorlinks=true, citecolor=blue, linkcolor=black, urlcolor=black]{hyperref}
\usepackage{orcidlink}
\usepackage{bm}
\usepackage{enumitem}

\begin{document}
\title{Photon arrival time lags in 3D radiative transfer with \texttt{SKIRT}}

\author{
Y.~Mochizuki\inst{1,2}\email{mochizuki-yuto652@g.ecc.u-tokyo.ac.jp}$^{\orcidlink{0000-0003-3224-1821}}$\fnmsep\thanks{mochizuki-yuto652@g.ecc.u-tokyo.ac.jp}
\and 
B.~Vander~Meulen\email{bert.vandermeulen@esa.int}\inst{3,4,5}$^{\orcidlink{0000-0002-5488-1961}}$
\and
M.~Tsujimoto\email{tsujimoto.masahiro@jaxa.jp}\inst{1,2}$^{\orcidlink{0000-0002-9184-5556}}$\fnmsep\thanks{tsujimoto.masahiro@jaxa.jp}
\and 
\\
P.~Camps\email{peter.camps@me.com}\inst{4}$^{\orcidlink{0000-0002-4479-4119}}$
\and 
M.~Baes\inst{4}\email{maarten.baes@ugent.be}$^{\orcidlink{0000-0002-3930-2757}}$
\and 
A.~Lauwers\inst{4}\email{arno.lauwers@ugent.be}$^{\orcidlink{0009-0003-9692-9382}}$
\and 
G.~Vanoppen\inst{6}\email{gijs.vanoppen@uhasselt.be}$^{\orcidlink{0009-0007-4399-5107}}$
}
\institute{
Institute of Space and Astronautical Science, Japan Aerospace Exploration Agency, 3-1-1
Yoshino-dai, Chuo-ku, Sagamihara, Kanagawa 252-5210, Japan
\and Department of Astronomy, Graduate School of Science, The University of Tokyo, 7-3-1
Hongo, Bunkyo-ku, Tokyo 113-0033, Japan
\and European Space Agency, European Space Research and Technology Centre, Keplerlaan 1,
2201 AZ Noordwijk, The Netherlands
\and Department of Physics and Astronomy, Universiteit Gent, Proeftuinstraat 86 N3, B-9000 Ghent, Belgium
\and SRON Netherlands Institute for Space Research, Niels Bohrweg 4, 2333 CA Leiden, The
Netherlands
\and Faculty of Sciences, Universiteit Hasselt, Martelarenlaan 42, 3500 Hasselt, Belgium
}
\date{Received May 6, 2026; accepted November 1, 2026}

\abstract
{Reprocessed X-ray emission from cold material surrounding compact objects carries
critical information about the geometry and physical state of the system. In particular,
the Fe~K$\alpha$ fluorescence line provides a powerful diagnostic through its spectral
and timing properties relative to the direct emission. Interpreting these observables
requires self-consistent modelling of radiative transfer (RT) in three-dimensional (3D)
space.}
{We aim to implement photon arrival time lags in \texttt{SKIRT}, a state-of-the-art
numerical RT code based on a Monte Carlo solver, thereby enabling self-consistent
modelling of both the spectral and timing properties of reprocessed emission in 3D
space.}
{We modified the code to accumulate photon path lengths over their lifetimes and convert
them into arrival time lags relative to the direct emission. The implementation is
validated using simple geometries and then applied to a realistic model of an active
galactic nucleus (AGN), for which we compute the time-lag and energy distribution and
the cross-correlation functions (CCFs) between the fluorescence line and broad bands.}
{The simulations reproduce the expected time-lag distributions and their energy
dependence for both free-electron scattering and the photoelectric absorption $+$
fluorescence. In the AGN application, the model naturally yields distinct lag components
from the inner and outer structures, as well as energy-dependent time lags associated
with the fluorescence line. The derived CCFs successfully recover both the peak time lag
and its finite width, which arise from the extended geometry and intrinsic variability.}
{The new timing capability enables self-consistent modelling of both the spectral and
timing properties of reprocessed emission, not only in X-rays but also at other
wavelengths. In X-rays, it provides a robust framework for interpreting
high-resolution spectroscopic data in conjunction with timing information from current
and future missions such as XRISM and NewAthena. A wide range of applications is
foreseen, including reverberation mapping studies of accreting systems.}


\maketitle
\nolinenumbers

\section{Introduction}\label{S1}

Compact objects (black holes, neutron stars, and white dwarfs) are often surrounded by
cold material, such as accretion disks, broad-line regions (BLRs), and dusty tori in
active galactic nuclei (AGNs) \citep[e.g.][]{antonucci1993,urry1995,netzer2015}; stellar
winds in high-mass X-ray binaries \citep[e.g.][]{macgregor1982}; and accretion streams
in cataclysmic variables and low-mass X-ray binaries \citep[e.g.][]{armitage1996}. The
structures of these surrounding media are key to understanding fundamental astrophysical
questions, such as the evolution of compact objects and their feedback on their host
environments \citep[e.g.][]{tauris2006,magorrian1998,fabian2012}. In particular, the
properties of the reprocessed emission from the surrounding material relative to the
direct emission from the compact object provide a powerful probe of their structure.

Of particular importance in the reprocessed emission is the Fe~K$\alpha$ fluorescence
line at 6.4~keV, which is widely observed in many systems due to the high abundance of
Fe and its large fluorescence yield \citep[e.g.][]{torrejon2010,shu2010a}. The
properties of this line relative to the direct emission are studied using both
spectroscopic and timing methods. In the spectroscopic method, the line strength
relative to the continuum level of the direct emission, typically quantified by the
equivalent width (EW), constrains the geometry of the compact object, the surrounding
material, and the observer
\citep[e.g.][]{blandford1982,peterson1993,peterson2001,peterson2004,cackett2021}. In the
timing method, the time lag of the line band relative to the continuum band of the
direct emission, measured using the cross-correlation function (CCF), enables
reverberation mapping of the surrounding material
\citep[e.g.][]{miller2018,zoghbi2019,noda2023}.

In the spectroscopic method, significant advances have been made in recent years with
the advent of X-ray microcalorimeter spectroscopy using the Resolve instrument
\citep{kelley2025,ishisaki2025a} on board the X-Ray Imaging and Spectroscopy Mission
(XRISM; \citealt{tashiro2025}). With its unprecedented spectral resolution for the
Fe~K$\alpha$ line ($R \equiv E / \Delta E \sim 1400$; \citealt{porter2025}), Resolve can
distinguish narrow and broad line components \citep{collaboration2024,bogensberger2025},
constrain bulk and thermal velocities \citep{audard2025,audard2026a}, characterise the
Compton shoulder of the line \citep{bianchi2026,thexrismcollaboration2026,fujiwara2026},
and reveal atomic shifts of the line due to slightly ionised Fe \citep{nagai2026}. These
diagnostics provide constraints on the geometry and physical state of the surrounding
material that are not accessible from the EW alone.

High spectral resolution is also beneficial for the timing method due to the effect of
spectral dilution. The CCF is typically constructed between a narrow energy band centred
on the Fe~K$\alpha$ line and a neighbouring continuum band. The narrow band contains not
only the line emission but also the underlying continuum emission. As a result, the
measured time lag is diluted by this continuum contamination
\citep{miller2010,omama2023}, an effect that can be mitigated with higher spectral
resolution.

To fully exploit observational data, both the spectroscopic and timing properties of the
Fe~K$\alpha$ line relative to the direct emission need to be interpreted
self-consistently using the same physical models. This requires radiative transfer (RT)
calculations with timing capabilities in three-dimensional (3D) space. Among various
numerical RT solvers, the Monte Carlo (MC) solver is particularly well suited for this
purpose. In this solver, individual photons are tracked throughout the transfer medium
from emission to either absorption or escape from the system
\citep{steinacker2013,noebauer2019}, and thus the implementation of the timing
capability is much more straight-forward compared to two-stream solvers such as
\texttt{Cloudy} \citep{ferland1998}, \texttt{XSTAR} \citep{kallman2004a}, and
\texttt{SPEX} \citep{kaastra1996} commonly employed in X-ray spectroscopic studies.

\texttt{SKIRT} is a state-of-the-art MCRT code \citep{baes2003,camps2015,camps2020} that
features flexibility in the configuration of the geometry, radiation sources, transfer
media, spatial grids, and detectors in 3D space. A key advantage of \texttt{SKIRT} is
its implementation of a suite of MC acceleration techniques, including peel-off, forced
scattering, continuous absorption \citep{baes2011}, composite biasing \citep{baes2016}, and
explicit absorption \citep{baes2022}. Without these techniques,
synthesised spectra would suffer from severe MC noise at the high spectral resolution
required for comparison with XRISM observations. \texttt{SKIRT} already incorporates the
X-ray atomic processes necessary to model the Fe~K$\alpha$ spectra
\citep{vandermeulen2023,vandermeulen2024}. The only missing capability for our purpose
is the treatment of timing.

\medskip

In this paper, we present the enhancement of timing capabilities in \texttt{SKIRT}.  We
first describe the implementation in Sect.~\ref{S2} and verify it using a simple
geometry in Sect.~\ref{S3}. In Sect.~\ref{S4}, we demonstrate its application to
Fe~K$\alpha$ line reverberation mapping of an AGN. Finally, we discuss some limitations
and outlook in Sect.~\ref{S5} before summarising the main results in Sect.~\ref{S6}.

\section{Implementation}\label{S2}

\subsection{Physical processes}\label{S2-1}
The Fe~K$\alpha$ fluorescence line is produced as a result of photoelectric absorption.
X-ray photons with energies above the threshold ($7.112$~keV for neutral Fe) cause
inner-shell ionisation, leaving a vacancy in the K shell. The resulting excited state
can decay via multiple channels, one of which is radiative decay producing an
Fe~K$\alpha$ ($n=2 \rightarrow 1$) fluorescence photon. Because a single fluorescent
photon is produced per photoionisation event, the fluorescence process can be treated
analogously to a scattering process in the MC framework, with a probability given by the
fluorescence yield of 0.3 \citep{krause1979}. This process is very rapid, as indicated
by the Einstein $A$ coefficient of $\approx 10^{14}$~s$^{-1}$ for Fe
\citep{palmeri2003}.
In competition with this is scattering by free electrons.  The free electron scattering
cross-section is comparable to the photo-electric absorption cross-section at the Fe~K
edge for cosmic abundances, making them the two dominant processes producing reprocessed
emission in the X-rays.  We hereafter refer to them as photo-electric
absorption/fluorescence (PA/FL) and free-electron scattering (ES).

These processes are already implemented in \texttt{SKIRT} \citep{vandermeulen2023}. For
the PA/FL process, the photoelectric absorption cross-section is provided for elements
with atomic numbers $Z=1, 2, \dots, 30$ using the formulae of
\cite{verner1995,verner1996}. The fluorescence yield is taken from
\citet{perkins1991}. The phase function is assumed to be isotropic. The emergent
Fe~K$\alpha$ fluorescence line complex includes fine-structure transitions
\citep{vandermeulen2024}, which can be resolved with Resolve.
For the ES process, the energy-dependent Klein-Nishina cross-section and scattering
phase function \citep{klein1929} are used \citep{vandermeulen2023}.  These processes,
along with others such as bound-electron scattering and dust scattering, can be enabled
or disabled individually in simulations \citep{vandermeulen2023}. Features involving
multiple processes, such as the Compton shoulder of the Fe~K$\alpha$ line, can be
computed self-consistently. The M\"uller matrix for each process is also implemented
\citep{vandermeulen2024a}, enabling the modelling of X-ray polarisation spectra.

\subsection{Algorithm}\label{S2-2}
\begin{figure}
 \centering
 \includegraphics[width=\linewidth,clip]{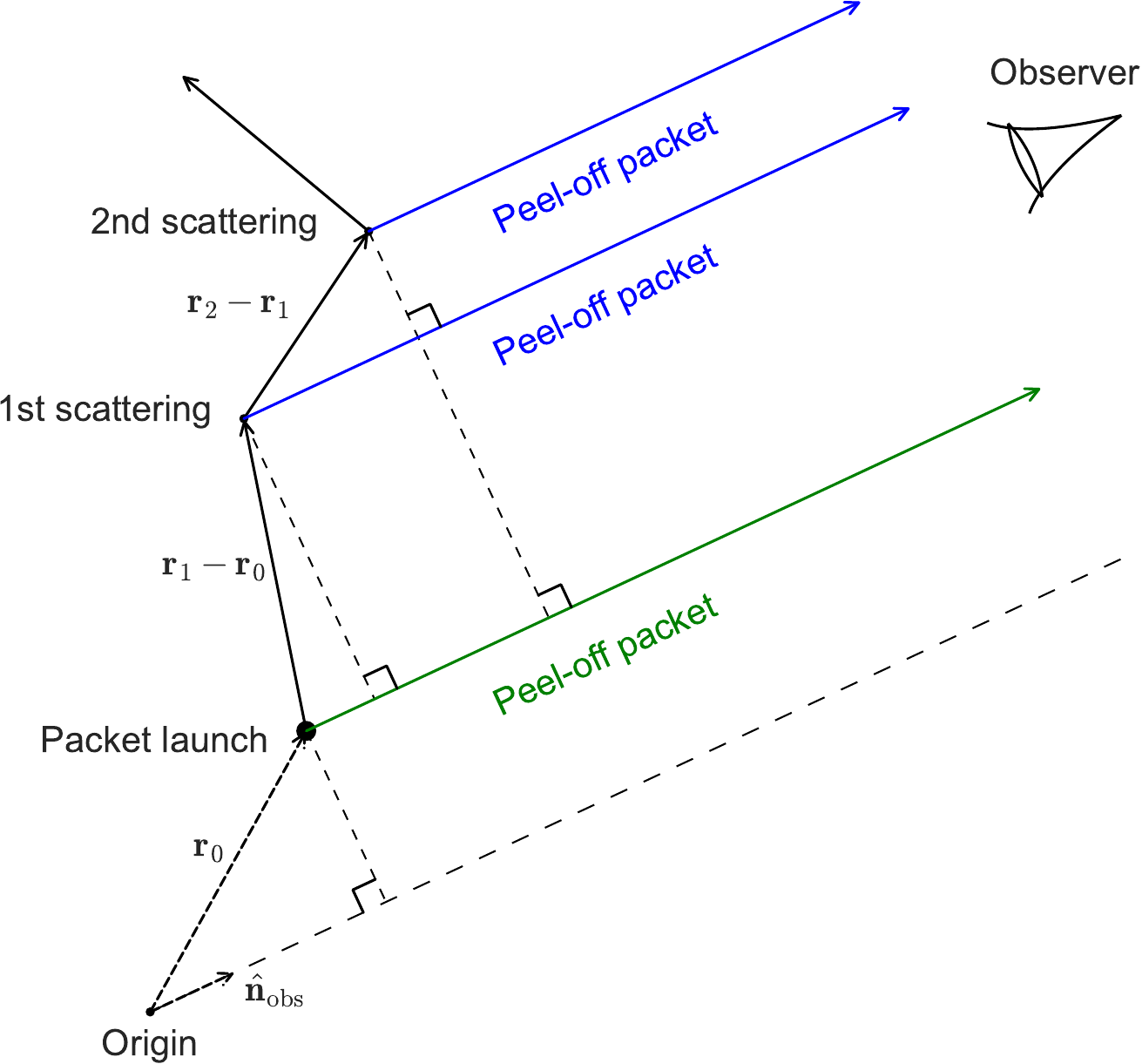}
 \caption{Concept of time tracking in \texttt{SKIRT}. A photon packet is launched from a
 point source located at position $\vec{r}_0$. Whenever a scattering event occurs at
 position $\vec{r}_i$ ($i = 1, 2, \ldots$), a peel-off photon is generated and directed
 towards the observer as reprocessed emission, while the original photon packet is
 scattered into a direction determined by the phase function. The observer detects the
 peel-off photons from both the source (green) and each scattering location (blue).}
\label{fig:implement}
\end{figure}

The algorithm for time tracking is straightforward. For each photon packet, we track the
path length, denoted as $\mathcal{D}$, covered from emission to either absorption or
escape from the system (Fig.~\ref{fig:implement}). We implement this in a manner
compatible with the peel-off technique \citep{Yusef-Zadeh1984}, which is a powerful
acceleration method for accumulating observed photon contributions.
At each scattering
and emission event, a peel-off photon is generated and directed towards the observer
placed outside the system. This approach significantly reduces the MC noise in the
reprocessed component of the synthesised spectra.

The origin of the arrival time is defined as the time at which a photon emitted at the
spatial origin arrives at the observer directly. 
The path length $\mathcal{D}$ is initialised to $0$ and calculated as the relative value with respect to this direct path. 
For peel-off photons emitted
from a source position ($\vec{r}_0$) in the simulation box,
the path length at emission is corrected to
\begin{equation}
 \label{eq:pp}
  \mathcal{D} \coloneqq - \vec{\hat{n}}_{\mathrm{obs}} \cdot \vec{r}_{0},
\end{equation}
where $\vec{\hat{n}}_{\mathrm{obs}}$ is the unit vector pointing towards the
observer. Therefore, the time lag can be negative for sources closer to the observer
than the spatial origin.

Let us assume that a photon experiences scattering at $\vec{r}_i$ ($i = 1, 2, \cdots$)
during its propagation. At each scattering event $i$, the path length is updated as
\begin{equation}
\label{eq:pp-update} 
 \mathcal{D} \coloneqq \mathcal{D} + |\vec{r}_i - \vec{r}_{i-1}|,
\end{equation}
to track for the distance travelled by the photon while scattering in the medium.  At
each scattering location, a peel-off photon is generated, for which the arrival time
correction is calculated as follows
\begin{equation}
\label{eq:ppp}
 \mathcal{D} \coloneqq \mathcal{D} - \vec{\hat{n}}_{\mathrm{obs}} \cdot \vec{r}_i,
\end{equation}
where the correction term represents the projected path length travelled in the
direction of the distant observer before escape (peel-off). 
For instance, the path length at the $N$-th scattering relative to direct
path is given by $\mathcal{D} = \sum_{i=1}^N |\vec{r}_i - \vec{r}_{i-1}| - \vec{\hat{n}}_{\mathrm{obs}} \cdot \vec{r}_N$.
The peel-off photons are
flagged with the number of scatterings experienced, allowing the synthesis of spectra
for a given value of $i$. Finally, the time lag is obtained by dividing the accumulated
path length relative to direct emission from the origin by the speed of light $c$. We
assume that no additional time lag is introduced by the scattering processes themselves.

The \texttt{SKIRT} code calculates the observed photon flux by integrating the weighted
flux contributions of peel-off photon packets in each arrival-time and energy bin. The
weights are calculated based on the probability that a photon actually propagates toward
the observer (Sect.~\ref{s5-3}). The resulting flux is expressed in units of photons
s$^{-1}$~cm$^{-2}$~keV$^{-1}$~s$^{-1}$. We refer to this as the `time-lag and energy
distribution', $F_\mathrm{t,E}(t,E)$. Both the arrival-time and energy bins can be
flexibly configured with either linear or logarithmic spacing.
The `time-lag distribution' is obtained by integrating over an energy range,
$F_\mathrm{t}(t) \equiv \int F_\mathrm{t,E}(t,E)\, \mathrm{d}E$, in units of photons
s$^{-1}$~cm$^{-2}$~s$^{-1}$. The `energy distribution', or energy spectrum, is obtained
by integrating over a time range, $F_\mathrm{E}(E) \equiv \int F_\mathrm{t,E}(t,E)\,
\mathrm{d}t$, in units of photons s$^{-1}$~cm$^{-2}$~keV$^{-1}$.

For the PA/FL process, we use the time-lag distribution obtained by integrating
$F_\mathrm{t,E}(t,E)$ over the 6.38--6.42~keV band to focus on the Fe~K$\alpha$ line
throughout the paper. For the ES process, we use the corresponding distribution obtained
by integrating over the designated range specific for the purpose of the verification and
demonstration.

\section{Verification}\label{S3}

\begin{figure}
 \centering
 \includegraphics[width=\linewidth,clip]{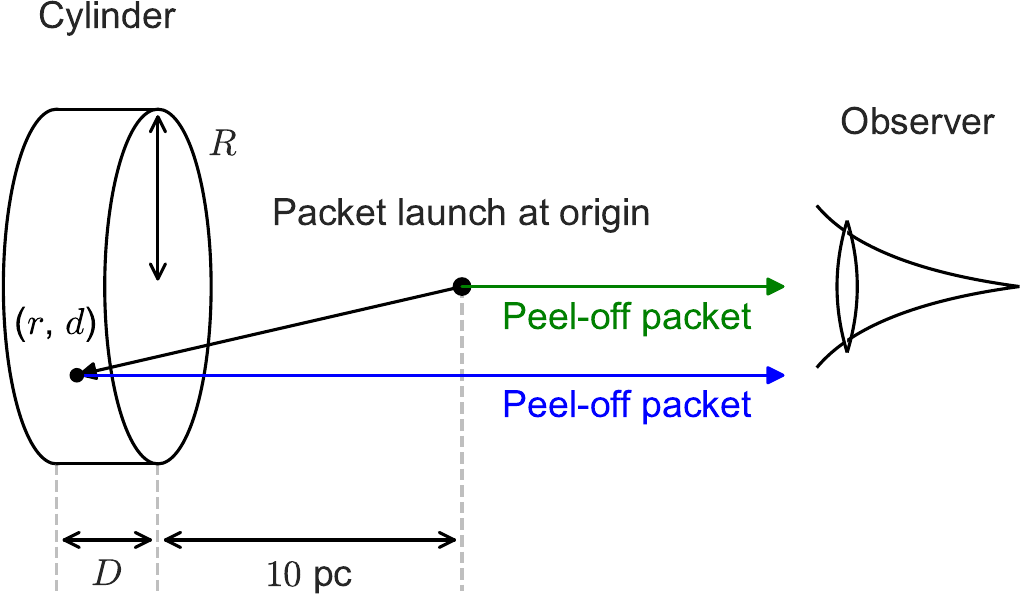}
\caption{Verification setup. Incident photons are emitted from a point source at the
origin. A cylinder is placed behind the source as seen from the observer. The time lag
between the direct and reprocessed emission at ($r$, $d$) is simulated.}
\label{fig:cylinder}
\end{figure}

We verify the timing implementation for PA/FL and ES separately, using the simple setup
shown in Fig.~\ref{fig:cylinder}. The incident source is located at the origin and
treated as an isotropic, unpolarised point source.  The reprocessed emission is produced
at a point $(r, d)$ within a cylinder located behind the source as seen by the observer,
where $r \in [0,R]$ is the radial distance from the central axis of the cylinder and $d
\in [0,D]$ is the depth from the front surface of the cylinder.  The geometry of the
cylinder is parameterised by its radius ($R$) and thickness ($D$). The distance from the
source to the cylinder is fixed at 10~pc, resulting in a minimum time lag of
$20~\mathrm{pc}/c$.

The cylinder is filled with a neutral gas with no dust, characterised by the uniform
hydrogen number density $n_{\mathrm{H}}$ in the case of PA/FL, or with a free-electron
medium characterised by the uniform electron number density $n_{\mathrm{e}}$ in the case
of ES. The corresponding hydrogen column density is $N_{\mathrm{H}} = n_{\mathrm{H}} D$
and the electron column density is $N_{\mathrm{e}} = n_{\mathrm{e}} D$. The Fe abundance
relative to H is denoted by $A_{\mathrm{Fe/H}}$. The material in the cylinder is assumed
to be at rest, with no thermal, bulk, or microturbulent motion.

We use two different optical depths as proxies for $N_{\mathrm{H}}$ and
$N_{\mathrm{e}}$. The first is for the PA/FL process. Assuming that the photoelectric
absorption cross-section at the K edge is $\sigma_{\mathrm{FeK}} = 3.36 \times
10^{-20}$~cm$^{2}$ \citep{verner1995,verner1996}, the corresponding optical depth is
$\tau_{\mathrm{FeK}} = A_{\mathrm{Fe/H}}\,\sigma_{\mathrm{FeK}}\,N_{\mathrm{H}}$.
The second is for the ES process. Assuming that the cross-section is
$\sigma_{\mathrm{es}}$, the corresponding optical depth is $\tau_{\mathrm{es}} =
\sigma_{\mathrm{es}}\,N_{\mathrm{e}}$, We adopt $\sigma_{\mathrm{es}} = 6.49 \times
10^{-25}$~cm$^{2}$ at 6.4~keV.

\medskip

We calculate the time-lag and energy distribution for the reprocessed emission. The
incident emission is assumed to have a power-law spectrum in energy with a photon index
of $\Gamma = 1$. The total luminosity is $5\times10^{41}$~erg~s$^{-1}$ in the 2--12~keV
band. The observer is placed at a distance of 10~Mpc from the source. Accordingly, the
time-lag and energy distribution of the incident emission is expressed as
$F_\mathrm{t,E}(t,E) \propto \delta(t)\,E^{-\Gamma}$, where $\delta(t)$ is the Dirac
delta function.  All simulations were run with $10^{9}$ photon packets.

We first verify whether the time-lag distribution, $F_\mathrm{t}(t)$, is consistent with
the cylinder geometry (Sect.~\ref{S3-1}). For simplicity, we include only the ES process
in the cylinder. We then examine the $F_\mathrm{t}(t)$ for two different processes
(Sect.~\ref{S3-2}) in preparation for the demonstration (Sect.~\ref{S4}). In this case, both the PA/FL and ES processes are tested. Several
combinations of the cylinder parameters ($r$, $d$, $\tau$, and $A_{\mathrm{Fe/H}}$) are
used for comparison.

\subsection{Time lag distributions}\label{S3-1}
The time-lag distribution is expected to reflect the spatial distribution of the
locations where reprocessing occurs in the cylinder. Based on single scattering in the
geometry, the time lag for single-scattered photons at radius $r$ and depth $d$ is given
by
\begin{equation}
 \label{e02}
  \delta t(r,d) \equiv \frac{\sqrt{(10~\mathrm{pc}+d)^{2}+r^{2}} + 10~\mathrm{pc} + d}{c}.
\end{equation}
The time lag is minimal at $(r=0,\,d=0)$ and maximal at $(r=R,d=D)$. This expectation is
modified by multiple scattering in the cylinder. We fix $R = 10$~pc and vary $D \in
\{10^{-9}, 10\}$~pc and $\tau_{\mathrm{es}} \in \{10^{-3}, 10^{-2}, 10^{-1}, 1,
10\}$. We used 2--12~keV band for evaluation.

\begin{figure}
\centering
 \includegraphics[width=\linewidth,clip]{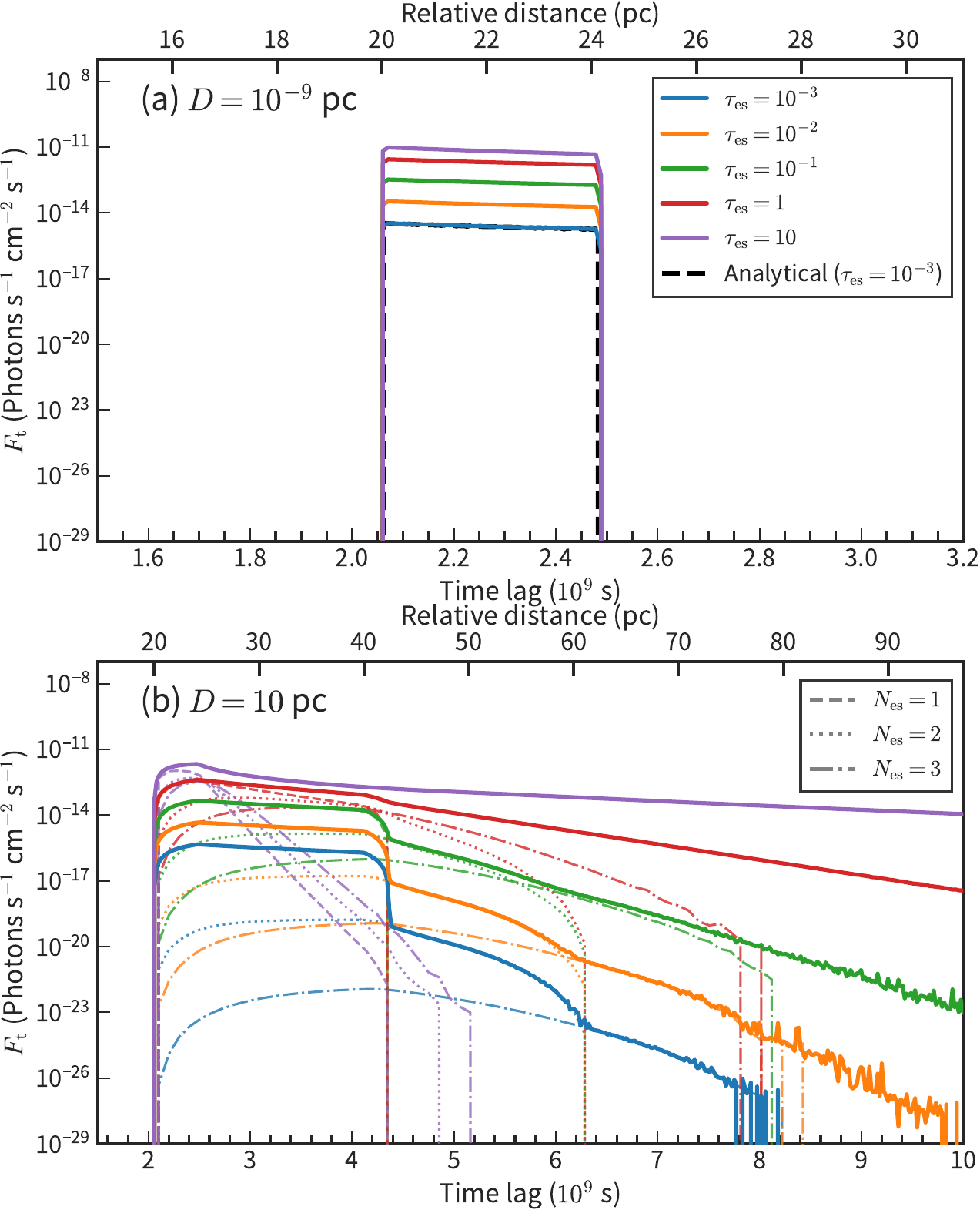}
 \caption{Time-lag distributions calculated with \texttt{SKIRT} for (a) a geometrically
 thin ($D = 10^{-9}$~pc) and (b) a geometrically thick ($D = 10$~pc) cylinder. In each
 case, results for five optical depths, $\tau_{\mathrm{es}} \in \{10^{-3}, 10^{-2},
 10^{-1}, 1, 10\}$, are shown in different colours. In panel (a), the analytical solution
 for single scattering is also shown for comparison. In panel (b), the distributions of
 single ($N_{\mathrm{es}} = 1$), double ($N_{\mathrm{es}} = 2$), and triple scattering
 ($N_{\mathrm{es}} = 3$) photons are shown separately in their respective colours.}
\label{fig:geometrical}
\end{figure}

Fig.~\ref{fig:geometrical} shows the time-lag distribution for (a) $D = 10^{-9}$~pc and
(b) $D = 10$~pc. In panel (a), where the cylinder is geometrically thin, the time-lag
distribution is primarily determined by the distribution of $r \in [0,R]$ in $\delta
t(r,d)$. The minimum and maximum time lags are consistent with $\delta t(r=0,d=0) =
20.0$~pc/$c$ and $\delta t(r=R,d=D) = 24.1$~pc/$c$, respectively. This is
particularly the case for $\tau_{\mathrm{es}} = 10^{-3}$. The simulated time-lag
distribution agrees well with the analytical solution considering single scattering
because multiple scattering is relatively weak at such low scattering optical
depths. The overall increase in the amplitude of $F_\mathrm{t}(t)$ from
$\tau_{\mathrm{es}} = 10^{-3}$ to 10 is due to the increase in the number of target
electrons in the cylinder.

In panel (b), the cylinder has a significant depth, and thus the effect of multiple
scattering is expected to be more apparent because a larger fraction of photons travel
over longer path lengths. We plot the time-lag distribution for single-, double-, and
triple-scattering photons separately, in addition to the total scattered photons. For
single scattering, the time-lag distribution is expected to lie between the minimum
$\delta t(r=0, d=0) = 20.0$~pc/$c$ and the maximum $\delta t(r=R, d=D) = 42.4$~pc/$c$.
As the number of scatterings increases, this distribution is progressively modified, but
with a decreasing contribution. For $\tau < 1$, the contribution of the
$N_{\mathrm{es}}$-th scattering scales approximately as $\tau^{N_{\mathrm{es}}}$. For
$\tau_{\mathrm{es}} = 10^{-3}$, including up to $N_{\mathrm{es}} = 3$ is sufficient to
reproduce the total scattered photons within the dynamic range of $F_{\mathrm{t}}(t)$
shown. For $\tau_{\mathrm{es}} = 1, 10$, higher orders of scattering contribute
significantly, and the resulting time-lag distribution exhibits a continuous smooth
tail.

\subsection{Different physical processes}\label{S3-2}
We next verify the implementation of the physical processes by comparing the result of
the two processes. In the time-lag and energy distribution $F_\mathrm{t,E}(t,E)$, one
process contributes more than the other at different energies. At $E = 6.4$~keV, the
PA/FL process dominates, whereas in the neighbouring continuum the ES process
dominates. We ran two cases: (i) only the ES process is included (as in
Sect.~\ref{S3-1}), and (ii) only the PA/FL process is included in the cylinder. We then
investigate the optical depth dependence for each. In both cases, we used the
6.38--6.42~keV band.

The radiation from the point source is assumed to be unpolarised and a collimated beam
directed towards the centre of the cylinder to make the geometrical interpretation
simpler.  We set the cylinder geometry to $R = 10^{-9}$~pc and $D = 10$~pc, effectively
treating it as a one-dimensional tube. We adopt $N_{\mathrm{H}} = 1.5 \times
10^{25}$~cm$^{-2}$ and $N_{\mathrm{e}} = 1.5 \times 10^{25}$~cm$^{-2}$ The cylinder is
optically thick for both processes.  For cosmic abundances ($A_{\mathrm{Fe/H}} = 4.68
\times 10^{-5}$; \citealt{anders1989a}), $\tau_{\mathrm{FeK}}$ is of the same order as
$\tau_{\mathrm{es}}$. 
To visually separate the simulation results for the two processes in the figure,
we adopt an Fe abundance 10 times higher than the cosmic value. With this choice,
$\tau_{\mathrm{es}} = 10$ and $\tau_{\mathrm{FeK}} = 236$ for the adopted
$N_{\mathrm{H}}$.

\begin{figure}
\centering
\includegraphics[width=\linewidth,clip]{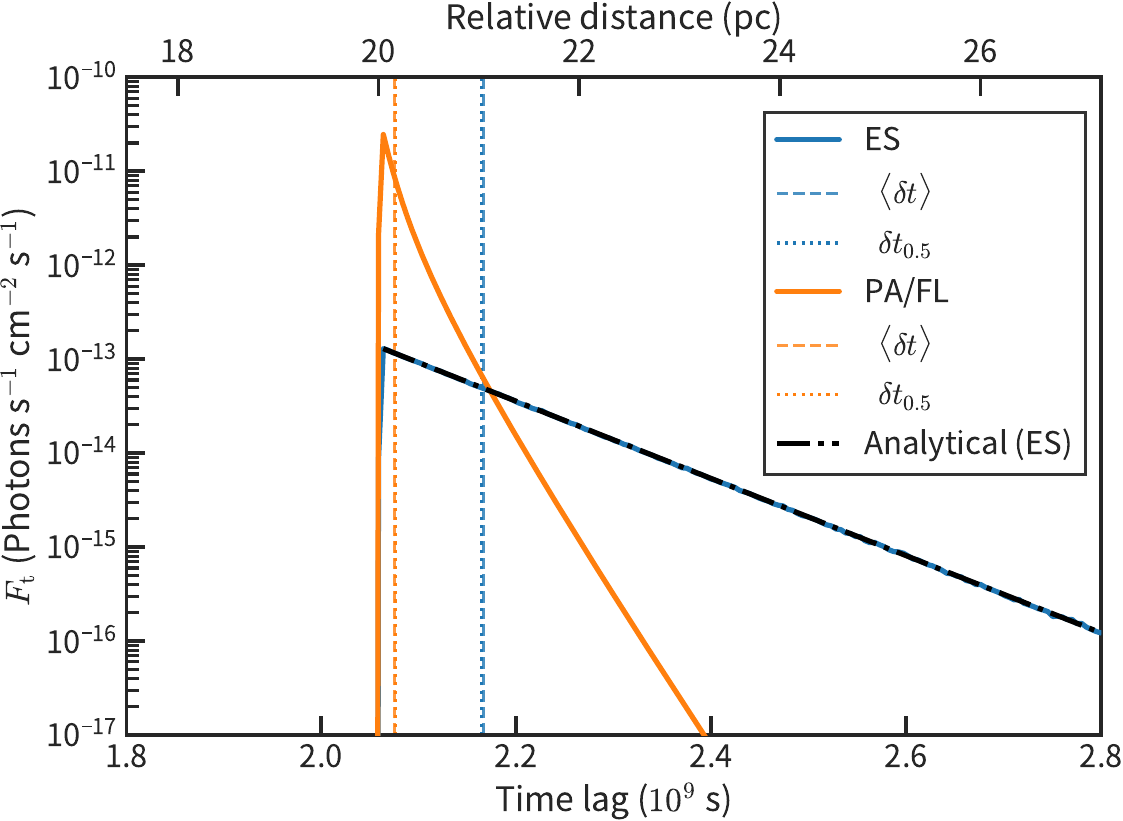}
 \caption{Time-lag distributions for free-electron scattering (ES; blue) and
 photoelectric absorption/fluorescence (PA/FL; orange), considered separately in a
 geometrically and optically thick cylinder. The Fe abundance relative to H is increased
 by a factor of 10 compared to the cosmic value to emphasise the differences. The
 average of each distribution (Eq.~(\ref{e03})) is indicated by dashed lines, while the
 geometrical time lag (Eq.~(\ref{e04})) is indicated by dotted lines. The analytical
 solution for the flux of the ES distribution is also plotted in black for comparison.}
\label{fig:physical}
\end{figure}

The results are shown in Fig.~\ref{fig:physical}. The minimum time lag is
$t_{\mathrm{min}} = 20$~pc$/c$, which holds for both processes. Because the optical
depth is large, there is no clear distinction between different scattering levels, and
the time-lag distribution roughly resembles exponential (straight line in log-linear plot), as in the case of ($D = 10$~pc, $\tau_{\mathrm{es}} = 10$) in Sect.~\ref{S3-1}.

The tail slope differs between the PA/FL and ES processes, as the characteristic time
lag is determined by the time required to reach the skin depth of the cylinder. The
PA/FL process has a smaller skin depth, and thus its time-lag distribution declines more
rapidly than that of the ES process. To quantify this, we calculate the average time lag
as
\begin{equation}
 \label{e03}
\langle \delta t \rangle \equiv \frac{\int_{t_{\mathrm{min}}}^{\infty} (t - t_{\mathrm{min}})\, F_\mathrm{t}(t)\, dt}{\int_{t_{\mathrm{min}}}^{\infty} F_\mathrm{t}(t)\, dt},
\end{equation}
and compare it with the geometrical time lag defined as
\begin{equation}
 \label{e04}
\delta t_{0.5} \equiv \frac{2 d_{0.5}}{c},
\end{equation}
where $d_{0.5}$ is the skin depth at which $\tau = 0.5$ is reached from the surface. 
The detected photons propagate from the source toward the cylinder and subsequently
escape toward the observer, meaning that the mean scattering point should occur at $\tau
= 0.5$.
For both the ES and PA/FL processes, the two quantities are in good agreement
(Fig.~\ref{fig:physical}), supporting our implementation and interpretation of the
results. A more quantitative estimate can be made for ES. It is expected to follow the
transmission probability through the cylinder. Given the cylinder thickness $D$ and the
optical depth $\tau_{\mathrm{es}}$, the intensity of photons travelling an additional
distance $2d = c \, \delta t$ is proportional to $\exp(-2\tau_{\mathrm{es}}d/D)$. Thus,
the time-lag distribution should follow
\begin{equation}
F_\mathrm{t}(t) \propto \exp\left(-\frac{\tau_{\mathrm{es}} c \, \delta t}{D}\right).
\label{e05}
\end{equation}
This analytical expectation is in good agreement with our numerical results, further
validating our implementation.

\section{Demonstration}\label{S4}

We confirmed that the time lag is calculated as expected and both the ES and PA/FL
processes are implemented as intended (Sect.~\ref{S3}). We now demonstrate the
capability of the code by applying it 
to an idealised AGN geometry designed to make the time lags clearly visible.
We derive the
energy-dependent time lag (Sect.~\ref{S4-1}) and the CCF for reverberation mapping
(Sect.~\ref{S4-2}).

\subsection{Energy-dependent time lag}\label{S4-1}
\subsubsection{Setup}\label{S4-1-1}
\begin{figure*}
\centering
\includegraphics[width=\linewidth,clip]{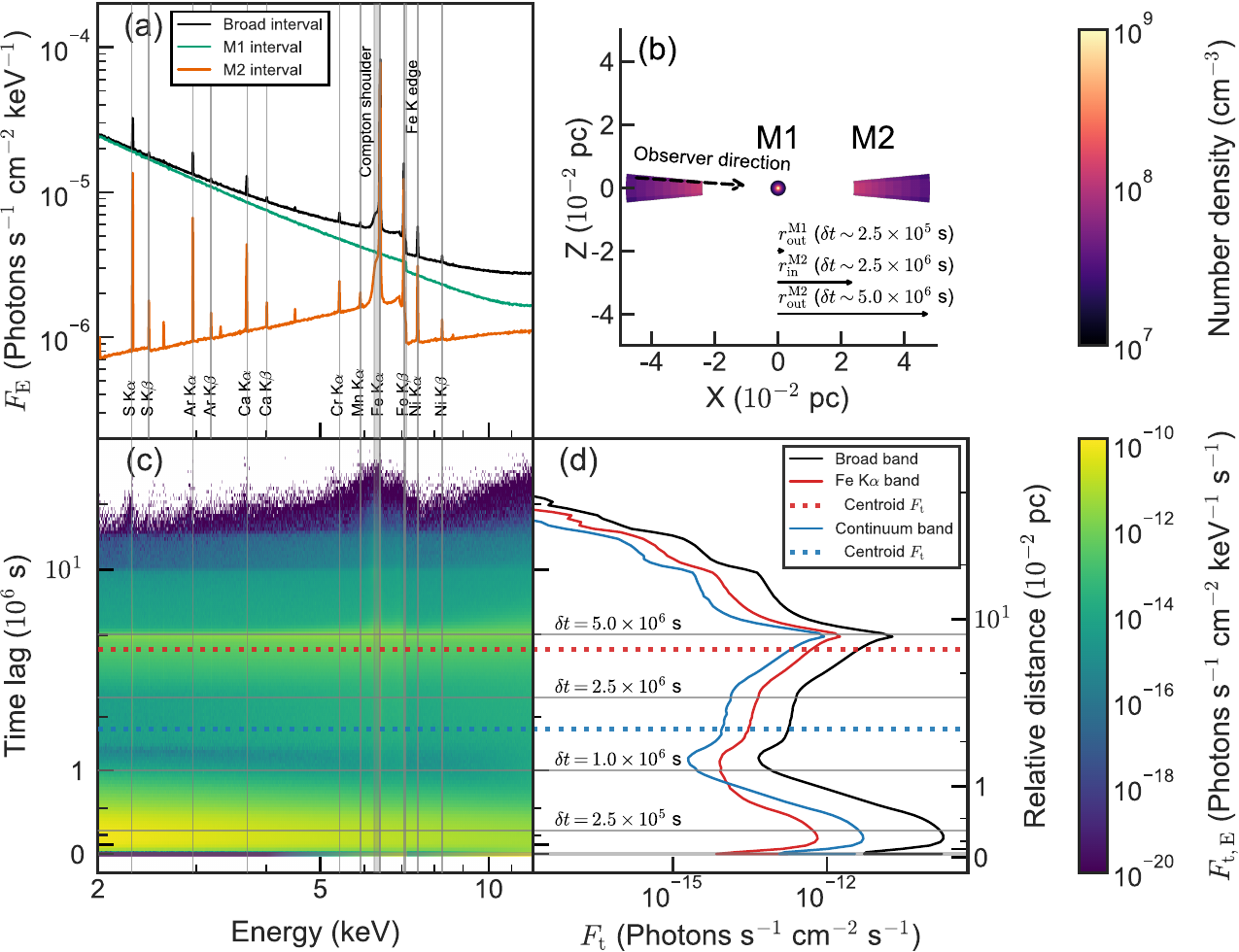}
 \caption{Simulation setup and results. Panel (b) shows the density map of the system,
 consisting of the incident source at the centre, a spherical fully ionised corona (M1),
 and a neutral torus (M2). Panel (c) displays the two-dimensional time-lag and energy
 distribution. The projections of this distribution onto the energy axis (energy
 spectrum) and the time axis (time-lag distribution) are shown in panels (a) and (d),
 respectively.  In panel (a), the spectra integrated over the broad time interval ($[0,
 10^8]$~s), M1 interval ($(0, 10^6]$~s), and M2 interval ($(10^6, 10^8]$~s) are shown,
 with prominent spectral features labelled.  In panel (d), the time-lag distributions for
 the broad energy band (2--12~keV), Fe~K$\alpha$ band (6.38--6.42~keV), and continuum
 band (6.50--6.90~keV) are shown. The centroid fluxes for the Fe~K$\alpha$ and
 continuum bands are indicated by dotted lines. The direct emission is represented by
 the grey band at $\delta t = 0$, while the secondary axis shows the corresponding
 geometrical distances.}
 \label{fig:AGN}
\end{figure*}

The setup is shown in Fig.~\ref{fig:AGN}~(b). The incident emission is assumed to
originate from a point source at the origin, representing a supermassive black hole
(SMBH) with a mass of $M_\bullet = 10^{7} M_\odot$. The spectrum is described by a power
law in energy with a photon index of $\Gamma = 1.8$ and a luminosity of
$5\times10^{41}$~erg~s$^{-1}$ in the 2--12~keV range. The radiation from the point
source is assumed to be unpolarised and isotropic.

Two distinct components, labelled M1 and M2, are placed around the incident source. In
both M1 and M2, the material has no velocity components (i.e. bulk motion, thermal
motion, and microturbulence are not considered). We adopt the cosmic abundances given by
\cite{anders1989a}. For both components, the line-of-sight column density is set to
$N_{\mathrm{e}} = 10^{25}$~cm$^{-2}$ in M1 and $N_{\mathrm{H}} = 10^{25}$~cm$^{-2}$ in
M2, and the number density decreases radially as $\propto r^{-2}$, where $r$ is the
distance from the incident source.

M1 is located in the vicinity of the SMBH, representing the corona. It is modelled as a
spherical gas distribution with an inner radius $r_{\mathrm{in}}^{\mathrm{M1}} =
10^{2}\, r_g = 4.8 \times 10^{-5}$~pc (equivalent to a time lag of $\delta t = 4.9
\times 10^3$~s) and an outer radius $r_{\mathrm{out}}^{\mathrm{M1}} = 5 \times 10^{3}\,
r_g = 2.4 \times 10^{-3}$~pc ($\delta t = 2.5 \times 10^5$~s), where $r_g \equiv
GM_\bullet/c^{2}$ is the gravitational radius of the SMBH. The material in M1 is assumed
to be fully ionised, so only the ES process operates.

M2 surrounds M1 and represents the torus and the BLR. It is modelled as an axisymmetric
torus with a half-opening angle of $5^{\circ}$. The inner
($r_{\mathrm{in}}^{\mathrm{M2}}$) and outer ($r_{\mathrm{out}}^{\mathrm{M2}}$) radii are
set to $5 \times 10^{4}\, r_g$ and $10^{5}\, r_g$, corresponding to physical distances
of $2.4 \times 10^{-2}$~pc ($\delta t \sim 2.5 \times 10^6$~s) and $4.8 \times
10^{-2}$~pc ($\delta t \sim 5.0 \times 10^6$~s), respectively. The material in M2 is
neutral, allowing both the ES and PA/FL processes to operate.

The time-lag distribution is divided into $10^{6}$ linearly spaced bins, corresponding
to light-travel distances of 1~pc. The energy spectrum is divided into 1000
logarithmically spaced bins from 2 to 12~keV, approximately matching the spectral
resolution of the Resolve instrument \citep{porter2025}.

The simulation was run with $10^{9}$ photon packets using JAXA's high performance
computer (JSS3; TOKI-RURI nodes, Intel Xeon Gold CPUs) as a single process using $24$ threads, consuming a total of $118$ core-hours. Enabling the newly implemented feature
to track photon arrival times resulted in a negligible ($\sim$ 1\%) increase in computational time compared
to the run without this feature.

\subsubsection{Results}\label{S4-1-2}
Fig.~\ref{fig:AGN}~(c) shows the time-lag and energy distribution, $F_\mathrm{t,E}(t,E)$, simulated
with \texttt{SKIRT} for a nearly edge-on inclination ($86^{\circ}$). Note that this
distribution includes both the reprocessed emission and the direct emission transmitted
through M1 and M2 with no time lag ($\delta t=0$). The projections of this distribution
onto the time axis (i.e. the time-lag distribution, $F_\mathrm{t}(t)$) and the energy axis
(i.e. the energy spectrum, $F_\mathrm{E}(E)$) are shown in Fig.~\ref{fig:AGN}~(d) and (a),
respectively. Both projections show distinct features. The features in the time-lag
distribution (Fig.~\ref{fig:AGN} (d)) correspond to geometrical distances divided by
$c$. The energy spectrum (Fig.~\ref{fig:AGN} (a)) reveals the K$\alpha$ and K$\beta$
fluorescence lines of major elements (S, Ar, Ca, Cr, Mn, Fe, and Ni), as well as the
Fe~K edge. The strongest line, Fe~K$\alpha$, exhibits a prominent Compton shoulder.

The time-lag and energy distribution (Fig.~\ref{fig:AGN} (c)) exhibits two major
components corresponding to the reprocessed emission from M1 and M2. They are separated
by a boundary at $\sim 10^{6}$~s, with shorter lags corresponding to M1 and longer lags
to M2.
In the M1 component, the time lag is nearly uniform across energies. This is because the
reprocessed emission is driven solely by the ES process, which is nearly
energy-independent in the response and reflect the spectral shape of the incident
spectrum, as shown in Fig.~\ref{fig:AGN} (a). This component peaks at
$r_{\mathrm{out}}^{\mathrm{M1}}/c \sim 2.5 \times 10^{5}$~s and dominates until $\delta
t \lesssim 10^{6}$~s due to multiple scattering in M1. Because the ES spectrum is nearly
flat within each neighbouring energy band, the band-limited flux contributed by M1 is
proportional to the width of the energy band. Consequently, the peak integrated flux in
the continuum band (6.50--6.90~keV) is higher than that in the Fe~K$\alpha$ line band
(6.38--6.42~keV) by an order of magnitude, as shown in Fig.~\ref{fig:AGN} (d).

\begin{figure}
\centering
\includegraphics[width=\linewidth,clip]{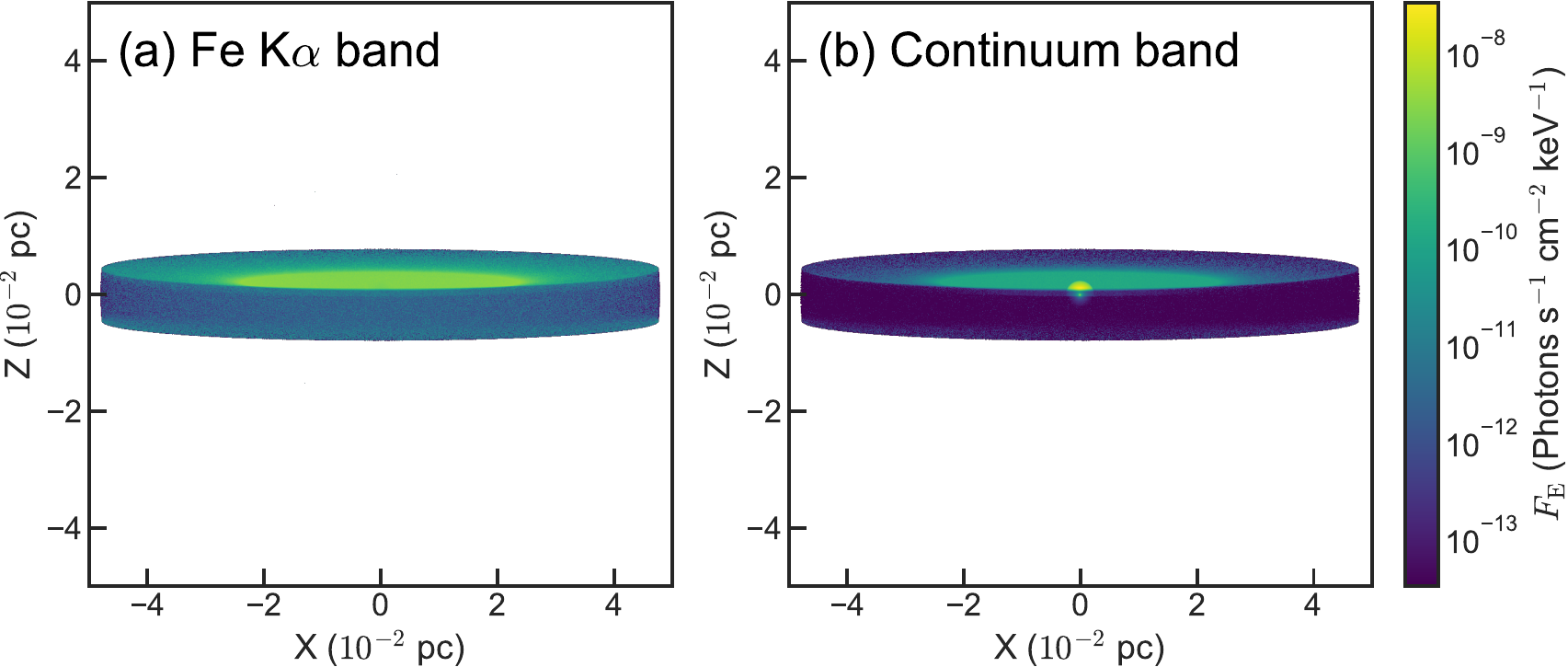}
 \caption{Synthesised image using the observed photons in the (a) Fe~K$\alpha$ band
 ($6.38$--$6.42$~keV) and (b) the continuum band ($6.50$--$6.90$~keV).}
 \label{fig:AGN-map}
\end{figure}

In contrast, the M2 component shows a strong energy dependence, as shown in
Fig.~\ref{fig:AGN} (a). It begins to contribute at the gap timescale ($\sim 10^{6}$~s)
and peaks at around $2 r_{\mathrm{in}}^{\mathrm{M2}}/c \sim 5 \times 10^{6}$~s, as the
observer primarily detects reprocessed emission at the inner edge of the torus on the
far side (Fig.~\ref{fig:AGN-map}). Around this peak, the Fe~K$\alpha$ band
(6.38--6.42~keV) shows a much higher flux than the neighbouring continuum band due to
the additional contribution from the PA/FL process. Beyond the peak, the time lag
continues to increase due to multiple scatterings. The overall trend of larger time lags
at higher energies arises because higher-energy photons can penetrate deeper into the
torus, reflecting the strong energy dependence of the photoelectric absorption
cross-section ($\sigma_{\mathrm{PA}}(E) \propto E^{-3}$). The time lag is locally
enhanced at the energies of the fluorescent lines, because these photons are produced by
higher-energy ionising photons that penetrate deeper into the torus. As a result, their
production sites are deeper than those of the adjacent continuum photons, leading to
longer total path lengths before escape.

\subsection{CCF}\label{S4-2}
\subsubsection{Setup}\label{S4-2-1}
\begin{figure}
\centering
\includegraphics[width=\linewidth,clip]{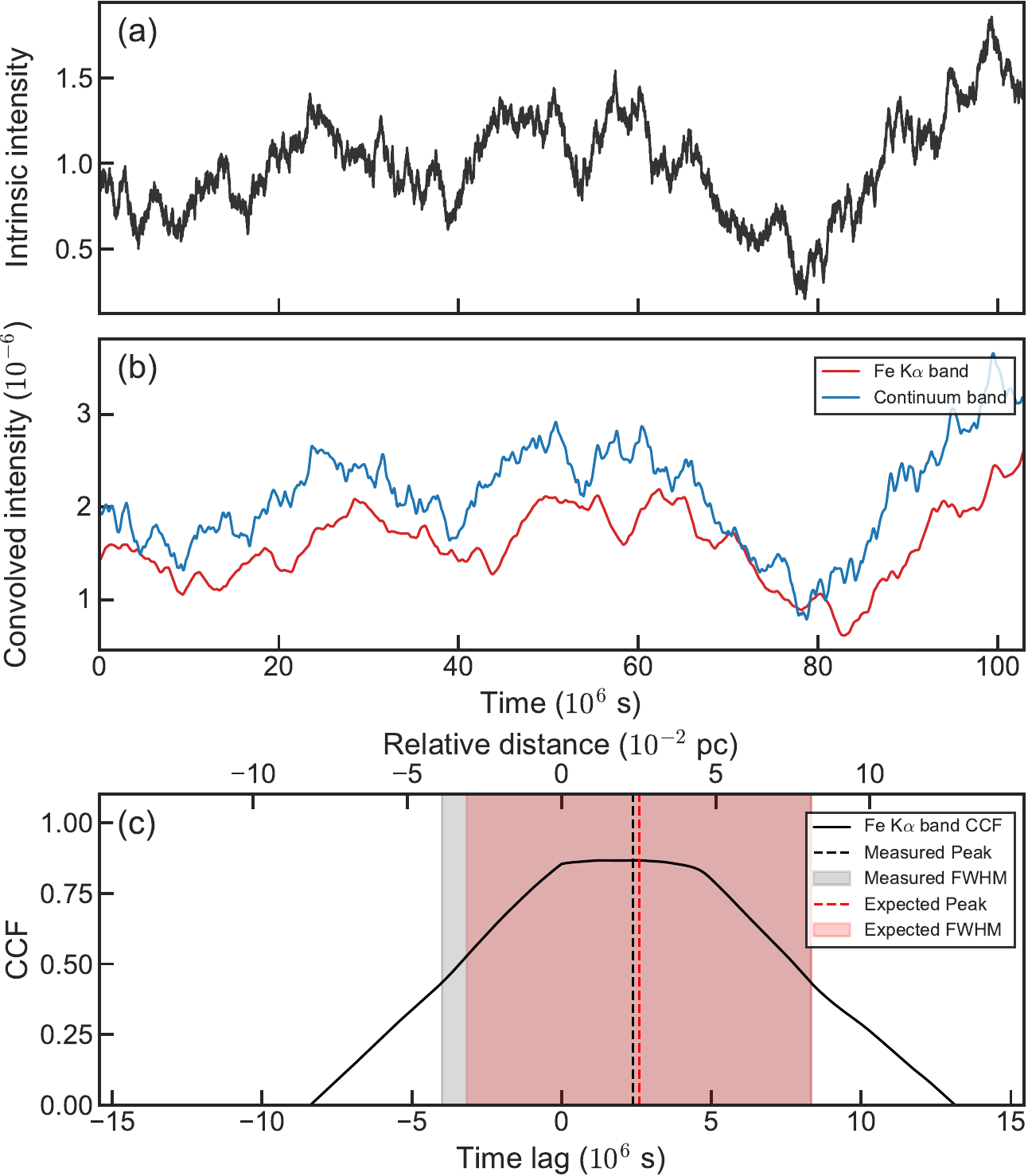}
 \caption{(a) Mock light curve of the incident emission, (b) simulated band-limited
 light curve of the Fe~K$\alpha$ band (6.38--6.42~keV; red) and the neighbouring
 continuum band (6.5--6.9~keV; blue), and (c) the CCF between the two light curves. The
 light curves are displayed only for $t>0$~s, but the actual calculation covers the
 margin before $t=0$. CCF is characterised by the peak and FWHM (black
 dashed line and shaded region), which are consistent with the expected values from
 calculation (red dotted line and shaded region).}
\label{fig:ccf}
\end{figure}

First, we generate a mock light curve for the incident source outside \texttt{SKIRT}.
We model the light curve using a damped random walk (DRW) process, which is widely used
to reproduce AGN variability \citep[e.g.][]{kelly2009}. The DRW process represents a
random walk with mean reversion; thus, its power spectral density (PSD) exhibits a white
noise profile ($\propto f^0$) at low frequencies and a brown noise profile ($\propto
f^{-2}$) at high frequencies, where $f$ is the frequency. We set the mean flux to 1 and
the standard deviation to 0.3.
The relaxation timescale, which characterises the return to the mean, is set to
108~days based on the mass-luminosity scaling relation \citep{mchardy2006}.
The resulting normalised mock light curve as a function of time $T$, $\tilde{I}(T)$, is shown in
Fig.~\ref{fig:ccf}~(a). We apply the same light curve across all energy bands.

The intensity of the incident emission at energy $E^{\prime}$ and time $T^{\prime}$ is expressed as
$F_0(T^{\prime},E^{\prime}) \propto \tilde{I}(T^{\prime})\, E^{\prime-\Gamma}$. The intensity of the observed emission (both
reprocessed and transmitted emission combined), $F_{\mathrm{obs}}(T,E)$, is obtained by
convolving this incident emission with the normalised time-lag and energy distribution prior to the energy redistribution integration,
$\tilde{F}_{\mathrm{T,T^{\prime},E,E^{\prime}}}(T - T^{\prime}, E, E^{\prime})$ as shown in Fig.~\ref{fig:AGN}~(c),
\begin{equation}
 \label{eq:convolution}
 F_{\mathrm{obs}}(T,E) = \int_{0}^{\infty} \mathrm{d}E^{\prime} \int_{-\infty}^{\infty} F_{0}(T^{\prime},E^{\prime})\, \tilde{F}_{\mathrm{T,T^{\prime},E,E^{\prime}}}(T - T^{\prime},E,E^{\prime})\, \mathrm{d}T^{\prime}.
\end{equation}
The result is then integrated over specific energy ranges to extract band-limited light
curves. For the light curve of the Fe~K$\alpha$ line,
\begin{equation}
 \label{eq:lc64}
  I_{\mathrm{line}}(T) \equiv \int_{6.38~\mathrm{keV}}^{6.42~\mathrm{keV}} F_{\mathrm{obs}}(T,E)\, \mathrm{d}E.
\end{equation}
For the light curve of the neighbouring continuum,
\begin{equation}
 \label{eq:lccont}
  I_{\mathrm{cont}}(T) \equiv \int_{6.50~\mathrm{keV}}^{6.90~\mathrm{keV}} F_{\mathrm{obs}}(T,E)\, \mathrm{d}E.
\end{equation}
These light curves are shown in Fig.~\ref{fig:ccf}~(b). Both curves are more smooth than
the incident light curve, as the spatial distribution of the reprocessing region
effectively acts as a low-pass filter for the reprocessed emission in the time
domain. The Fe~K$\alpha$ light curve is clearly delayed relative to the continuum light
curve.

We then compute the CCF between the two band-limited light curves. After
normalising both light curves, the CCF as a function of time lag $t$ is given by
\begin{equation}
 \label{eq:ccf}
 \mathrm{CCF}(t) \equiv \int_{-\infty}^{\infty}
 I_{\mathrm{line}}(T)\, I_{\mathrm{cont}}(T - t)\, \mathrm{d}T.
\end{equation}

\subsubsection{Results}\label{S4-2-2}
The resulting CCF is shown in Fig.~\ref{fig:ccf}~(c). First, the peak of the CCF,
which corresponds to the time lag of the Fe~K$\alpha$ band relative to the continuum band,
is $\sim 2.39 \times 10^6$~s (equivalent to $\sim 28$~days). This value is in good agreement with both the physical
distance between M1 and M2 (corresponding to a light-travel time of
$\sim 2.25 \times 10^6$~s) and the difference in the mean time lags derived from the
time-lag distribution (Fig.~\ref{fig:AGN}~(c)) for the Fe~K$\alpha$ and continuum band
emissions ($\sim 2.59 \times 10^6$~s).

Second, the CCF exhibits a finite width of $\sim 1.2 \times 10^{7}$~s around its peak.
This reflects the fact that the Fe~K$\alpha$ and continuum band emissions originate from
extended spatial regions. In addition, the autocorrelation function (ACF) of the
intrinsic light curve (Fig.~\ref{fig:ccf}~(a)) also contributes to the overall width of
the CCF. For comparison, we derive the width of the time response for the Fe~K$\alpha$
to be $\sim 3.77 \times 10^6$~s and continuum band emissions from the time-lag
distribution to be $\sim 5.13 \times 10^6$~s (Fig.~\ref{fig:AGN}~(d)) and the ACF width
from the mock light curve to be $\sim 9.63 \times 10^6$~s (Fig.~\ref{fig:ccf}~(a)), and
compute their quadrature sum as $\sim 1.15 \times 10^7$~s. This is in excellent
agreement with the simulated width of $\sim 1.23 \times 10^{7}$~s.

\section{Discussion}\label{S5}

The implementation presented in this work introduces a general framework to compute
photon arrival-time lags within the MCRT scheme of \texttt{SKIRT}. In this paper, the
methodology has been demonstrated for X-ray radiative transfer involving electron
scattering and photoelectric absorption followed by fluorescence. While the formalism
itself is broadly applicable, its interpretation relies on a number of physical and
numerical assumptions that define the regimes in which it can be robustly used.

\subsection{Applicability to different radiative processes}\label{s5-1}
The key assumption underlying the present implementation is that the time lag associated
with each photon is fully determined by its geometrical path length, implying that
interaction processes can be treated as effectively instantaneous. This is an excellent
approximation for the processes considered in this work. In particular, electron
scattering is intrinsically instantaneous, and fluorescence processes such as Fe
K$\alpha$ emission occur on timescales ($\sim 10^{-14}$\,s) that are entirely negligible
compared to the light-travel times relevant for the systems under study
\citep{palmeri2003}.

The same methodology and formalism are directly applicable to a broader class of
radiative transfer problems dominated by such instantaneous interactions. Examples
include resonance scattering and dust scattering, both of which are implemented in
\texttt{SKIRT} \citep{vandermeulen2023,camps2021,sameshima2026}. However, care must be
taken when extending the approach to processes in which energy exchange is not local in
time.

A clear example is dust radiative transfer.  Dust absorption followed by thermal
re-emission is often treated as an effective scattering event in MC-RT
simulations \citep{Bjorkman2001, Baes2005, Heymann2012}. However, unlike the
instantaneous scattering by dust grains, this process in reality involves energy storage
and re-emission over a finite timescale, which depends on grain properties and local
conditions.
A similar caveat may apply to photoionised plasmas, where recombination emission and
line emission processes can introduce additional time lags. Typical recombination
timescales depend strongly on density and ionisation state and can range from
effectively instantaneous in dense environments to dynamically relevant in more rarefied
media \citep{Osterbrock2006}. In such cases, the present implementation would not
capture the full time-delay behaviour, and a more general treatment including
process-dependent delay times would be required.

\subsection{Relativistic effects and geometry}\label{s5-2}
The current implementation assumes photon propagation in a Euclidean space, neglecting
general relativistic (GR) effects. This approximation is appropriate for regions
sufficiently far from compact objects, where spacetime curvature is weak. In the
demonstration presented in Sect.~\ref{S4}, the inner corona extends down to $\sim
10^{2}$~$r_g$. At these scales, relativistic effects may already introduce
non-negligible corrections to both the spectral and timing properties. In particular,
light bending can increase effective path lengths, while gravitational time dilation
alters the mapping between emission and observer time. Neglecting these effects may
therefore bias the inferred geometry and size scales of the emitting regions.

A full treatment of these regimes would require coupling the radiative transfer
calculation to a relativistic ray-tracing framework. Such an extension is currently not
foreseen within \texttt{SKIRT} itself. However, dedicated GR radiative transfer tools
are available and could, in principle, be combined with the present approach. Examples
include {\texttt{MONK}} \citep{Zhang2019} and {\texttt{RAIKOU}}
\citep{Kawashima2023}. Exploring such connections would be a promising direction for
future work.

\subsection{Compatibility with \texttt{SKIRT} features and Monte Carlo performance}
\label{s5-3}

A major strength of this work lies in implementing the timing capability within the established \texttt{SKIRT} framework.
\texttt{SKIRT} supports several physical effects that were not explored in this work but combine naturally with the timing framework. Doppler shifts due to gas kinematics are fully supported and can be incorporated consistently into the radiative transfer calculation \citep{BarrientosAcevedo2023}. Polarisation is likewise implemented \citep{peest2017,vandermeulen2024a} and can be combined with time-lag measurements, provided that suitable instruments are defined.

Besides peel-off (Sect.~\ref{S2-2}), the photon-packet life cycle in \texttt{SKIRT} employs several variance-reduction techniques that steer the MC process towards outcomes of interest \citep{steinacker2013,noebauer2019,camps2020}: at each event, the outcome is drawn from a biased rather than the physical probability distribution, and the packet weight, i.e.\ the fraction of the source luminosity carried by the packet, is adjusted to compensate. For example, the initial wavelength is by default drawn half of the time from the source spectrum and half of the time from a smooth logarithmic distribution, so that spectral ranges of low luminosity are also well sampled (composite biasing; \citealp{baes2016}). In forced scattering, a packet is not allowed to escape and instead its weight is lowered by the fraction that would have escaped, so that the same packet samples many scattering events until its weight becomes negligible \citep{baes2011}. At each emission and scattering event, a peel-off packet is sent to every instrument, with a weight equal to that of the parent packet multiplied by the emission profile or scattering phase function for the direction of the instrument, and attenuated by the extinction along the line of sight. The instrument adds this weighted contribution to the appropriate spatial, spectral, and time-lag bin.

Because of these successive biases, no simple formula describes the MC noise of the recorded results. \texttt{SKIRT} therefore offers optional statistics: for each bin of a spectrum, image, light curve, or spectral-time map, an instrument can record the sums of the first four powers of the packet contributions (the contribution moments). Contributions to a bin from the same launched packet are combined first, as they are not independent; these statistics cannot be reconstructed after rebinning. From these moments, \texttt{SKIRT} derives the relative error $R$ (the estimated standard error of a bin value relative to that value) and the variance of the variance (VOV), which quantifies the reliability of $R$ and is far more sensitive to rare, large contributions. Both are adopted from the nuclear particle transport code MCNP \citep{XMonteCarloTeam2003,camps2018}. Empirically, $R < 0.1$ and $\mathrm{VOV} < 0.1$ indicate a reliable result and allow $R$ to be interpreted as a relative error \citep{XMonteCarloTeam2003,camps2018, camps2022}. The value of $R$ therefore provides a direct check of whether the number of photon packets, $N_{\mathrm{pp}}$, is sufficient. 
High-precision time-lag distributions or higher-order timing diagnostics may demand a substantial $N_{\mathrm{pp}}$, particularly in optically thick or multiple-scattering regimes.

The efficiency of a simulation is quantified by the figure of merit (FOM),
\begin{equation}
  \mathrm{FOM} \equiv \frac{1}{R^{2}\,T},
  \label{eq:fom}
\end{equation}
where $T$ the computing time. For large $N_{\mathrm{pp}}$, $T \propto N_{\mathrm{pp}}$ and $R^{2} \propto 1/N_{\mathrm{pp}}$, so the FOM converges to a constant \citep{camps2018}. The FOM, evaluated per bin, thus allows the effectiveness of different techniques to be compared. For the timing feature, statistical efficiency alone is not sufficient: the results must also remain physically correct, which requires that the propagation distance of each sampled photon packet is tracked properly.

We assessed each acceleration technique for both correctness and efficiency. Source biasing (spatial and spectral), the absorption/scattering split, forced scattering, and continuous absorption \citep{baes2011} are fully compatible with the timing feature: they change how events are sampled and weighted, but not the bookkeeping of propagation distances. Continuous absorption is relevant only when the radiation field is stored, which is not required in the X-ray regime. Explicit absorption, intended for media with stimulated emission \citep{baes2022}, is also compatible but should be enabled only for such media. The one exception is line-core skipping, used in Ly$\alpha$ radiative transfer to bypass large numbers of resonant scattering events \citep{Ahn2002,Dijkstra2006,Smith2015}: the propagation distances and time delays associated with the skipped events are not tracked, so time-lag results must be interpreted with particular care.

One further setting deserves attention: forced-scattering packets are terminated once their weight has dropped below a fixed fraction ($10^{-4}$ by default) of the initial weight. As late-arriving flux is carried by long-lived, low-weight packets, this termination criterion truncates the tail of the time-lag distribution regardless of $N_{\mathrm{pp}}$. In a dusty slab of transverse optical depth 5 \citep[cf.][]{gordon2017}, lowering the threshold to $10^{-9}$ extended the reliable dynamic range of the light curve from about 5 to about 10 orders of magnitude below its peak.

Finally, path length stretching (PLS; \citealp{baes2016}) draws, for half of the events by default, the optical depth to the next interaction uniformly over the optical depth available along the path instead of from the physical exponential distribution. Packets thus reach high optical depths more often. PLS can significantly accelerate typical dusty-galaxy simulations, but is less suited to regimes of high optical depth or high albedo, where compounded bias factors produce rare packets with very large weights. It is also not compatible with large wavelength jumps in scattering events, such as those in our fluorescence implementation: in the X-ray torus benchmark of \citet{vandermeulen2023}, PLS reduced the FOM to two-thirds or less of its value without PLS. PLS is compatible with the timing feature, provided that the termination threshold is lowered as described above, since PLS lowers some packet weights even faster. For most X-ray applications, however, PLS is not advantageous because of the importance of fluorescence.

\subsection{Outlook}\label{s5-4}
Within these limitations, the present implementation provides a flexible framework for
modelling the coupled spectral and timing properties of radiation in complex
three-dimensional geometries. In addition to the Fe K$\alpha$ reverberation mapping used
in Sect.~\ref{S4}, it can be used for hard/soft-lag of the continuum emission as a
function of energy often seen in accreting systems \citep{fabian2009,mizumoto2019}, dust
scattering echo of SNR and X-ray binaries \citep{dwek2008,heinz2016},
and pulsar timing
delay due to scattering \citep{hickox2005}.
In particular, X-ray reflection from molecular clouds has been observed via Fe K$\alpha$ emission from the Sagittarius B2 region \citep{koyama1996,nobukawa2011} and through the apparent superluminal motion of the reflected emission around Sagittarius~A$^\ast$ \citep{ponti2010}, and it has also been extensively investigated using MCRT modelling \citep{sunyaev1998,odaka2011}.
These phenomena are actually observed with
contemporary observatories. With the future observing instruments such as the X-IFU
\citep{barret2018} onboard NewAthena, we will be able to construct the time-lag and
energy distribution like Fig.~\ref{fig:AGN} (c) directly from observations.

\section{Summary}\label{S6}

We have implemented the photon arrival time lag in the 3D MC-RT code
\texttt{SKIRT}. This framework calculates the time lag by accumulating the relative path
length of each photon packet and scaling it by the speed of light. We verified the
algorithm using a simple cylindrical geometry, confirming that the time-lag distribution
and its two distinct physical processes are reproduced as intended for both free-electron scattering
and photoelectric absorption/fluorescence processes.

We then demonstrate the capability of this feature by applying it to a realistic AGN
model.  Our results reproduce the expected CCF, explaining both the peak time lag and
its width of the distribution in terms of the extended spatial geometry and the
intrinsic light curve. These results show that photon arrival time lags have been
self-consistently and robustly incorporated into \texttt{SKIRT}.

\begin{acknowledgements}
 We thank the anonymous reviewer for the helpful comments and suggestions that improved this work.
 Y.\,M. acknowledges the hospitality by colleagues at ESTEC and Ghent University, where
 he stayed throughout the course of this work. This research made use of the JAXA's
 high-performance computing system JSS3. Y.\,M. acknowledges support from the JSPS
 Core-to-Core Program (grant number: JPJSCCA20220002), the JSPS KAKENHI (grant numbers:
 JP25KJ0923 and 26H02075). B.\,V.\ acknowledges support through the European Space
 Agency (ESA) Research Fellowship Programme in Space Science.
\end{acknowledgements}

\bibliographystyle{bibtex/aa}
\bibliography{bibtex/aa}

\end{document}